\documentclass{article}
\usepackage{spconf,amsmath,graphicx}
\usepackage{cite}
\usepackage[linesnumbered, ruled]{algorithm2e}
\usepackage{subfigure}
\usepackage{float}

\usepackage[utf8]{inputenc}
\usepackage[T1]{fontenc}
\usepackage[english]{babel}

\usepackage{cleveref}
\usepackage{autonum}

\let\oldnl\nl
\newcommand{\nonl}{\renewcommand{\nl}{\let\nl\oldnl}}

\DeclareMathOperator*{\argmin}{\arg\!\min}

\title{ROBUST CALIBRATION OF RADIO INTERFEROMETERS \\ IN MULTI-FREQUENCY SCENARIO}
\name{V. Ollier$^{\star}$ $^{\diamond}$ \qquad M. N. El Korso$^{\dagger}$ \qquad
A. Ferrari$^{\ddag}$ \qquad R. Boyer$^{\diamond}$ \qquad P. Larzabal$^{\star}$\thanks{This work was supported by: MAGELLAN (ANR-14-CE23-0004-01), ON FIRE project (Jeunes Chercheurs GDR-ISIS), ANR ASTRID project MARGARITA (ANR-17-ASTR-0015) and HyperFusion project.}}
\address{$^{\star}$ SATIE, UMR 8029, ENS Paris-Saclay, Cachan, France \\
$^{\diamond}$ L2S, UMR 8506, Universit\'{e} Paris-Sud,
Gif-sur-Yvette, France\\
$^{\dagger}$ LEME, EA 4416, Universit\'{e} Paris-Nanterre, Ville d'Avray, France \\
$^{\ddag}$ Laboratoire J.L. Lagrange, UMR 7293, Universit\'{e} Nice Sophia-Antipolis, France
}
\begin{document}
\ninept
\maketitle
\begin{abstract}

This paper investigates calibration of sensor arrays in the radio astronomy context. Current and future radio telescopes require computationally efficient algorithms to overcome the new technical challenges as large collecting area, wide field of view and huge data volume. Specifically, we study the calibration of radio interferometry stations with significant direction dependent distortions. We propose an iterative robust calibration algorithm based on a relaxed maximum likelihood estimator for a specific context: i) observations are affected by the presence of outliers and ii) parameters of interest have a specific structure depending on frequency. Variation of parameters across frequency is addressed through a distributed procedure, which is consistent with the new radio synthesis arrays where the full observing bandwidth is divided into multiple frequency channels. Numerical simulations reveal that the proposed robust distributed calibration estimator outperforms the conventional non-robust algorithm and/or the mono-frequency case.

\end{abstract}
\section{Introduction}
\label{sec:intro}

 With a resolution and sensitivity greater than any previous systems in the low observing frequencies, the low frequency array (LOFAR) and the square kilometre array (SKA) impose major challenges in terms of telescope design and data processing \cite{wijnholds2014signal,ferrariSKA,zarkaNENUFAR}. Among these challenges, the calibration step is crucial for the advanced phased array radio telescopes due to the large number of receivers and their wide field-of-view. These design features result in direction dependent effects \cite{jeffs2006direction}, varying over the field-of-view, due, \textit{e.g.}, to ionospheric disturbances, and in gain differences from one receiving element to another. To avoid any calibration error preventing the exploitation of the full sensitivity potential and decreasing the high dynamic range performance of imaging \cite{rau2009advances},  environmental and instrumental unknowns need to be corrected for.
 
In several studies \cite{boonstra2003gain,wijnholds2009multisource}, array signal processing tools are used to perform calibration, especially the maximum likelihood (ML) estimator which is usually introduced under a Gaussian noise model assumption \cite{yatawatta2009radio}. However, in our application, such noise modeling is not adapted: the sky model is composed of several bright known sources (the calibrator sources) but also many unknown unmodeled sources which give rise to incomplete sky models \cite{yatawatta2014robust}. Furthermore, radio astronomical measurements are contaminated by radio inteference \cite{raza2002spatial}. All these effects, and others, lead to the presence of outliers in the data set. To take them into account, a robust calibration technique  was introduced in \cite{yatawatta2014robust} where the noise model is specifically described as a Student's t with independent identically distributed entries. To improve calibration, we proposed, in a previous work \cite{ollier2016relaxed}, to adopt a compound-Gaussian modeling \cite{ollila2012complex} which includes a broad range of different distributions and revealed to be more robust \cite{ollierjournal}.

Furthermore, it should be noted that direction independent gains of each sensor and direction dependent perturbations associated to each source are estimated through the calibration process, where the latter parameters are frequency dependent. To exploit the known structure of variation \textit{w.r.t.} frequency \cite{martinjournalnew}, calibration is reformulated as a constrained consensus problem and addressed with the alternating direction method of multipliers (ADMM) \cite{bertsekas1989parallel,boyd2011distributed}. To reduce the computational burden, we process the data thanks to a distributed architecture with a network of agents \cite{chang2015multi}. Decentralized and distributed strategies have already been applied for image reconstruction in radio astronomy \cite{ferrari2014distributed} and also for calibration \cite{yatawatta2015distributed} in the Gaussian and non-structured case.
In this work, we particularly focus on robust calibration of radio interferometry  compact stations with direction dependent distortions, named as the 3DC calibration regime (structured case) \cite{ollierjournal}.

The notation used through this paper is the following: symbols $\left( \cdot \right) ^{T}$, $\left( \cdot \right)^{\ast }$, $\left( \cdot \right) ^{H}$ denote, respectively, the transpose, the complex conjugate and the Hermitian transpose. The symbol $\otimes$ represents the Kronecker product, $\mathrm{vec}(\cdot)$ stacks the columns of a matrix on top of one another, $\mathrm{diag}\{\cdot\}$ converts a vector into a diagonal matrix and
the trace is given by $\mathrm{tr}\left\{ \cdot \right\} $. The $B \times B$ identity matrix is referred by $\mathbf{I}_{B}$, $||\cdot||_F$ is the Frobenius norm, while $||\cdot||_2$ denotes the $l_2$ norm. Finally, $j$ is the complex number whose square equals $-1$ and $[\cdot]_k$ refers to the \textit{k}-th entry of the considered vector.

\section{Model set up and background on robust calibration algorithm}

\subsection{Data model}

We consider $D$ signal waves emitted by calibrator sources impinging on an array of $M$ antennas. The cross correlation between (noisy free) voltages measured by two antennas $p$ and $q$, is given by \cite{KerThesis2012}
\begin{equation}
\label{ME}
\mathbf{V}_{pq}(\boldsymbol{\theta}) =
\sum_{i=1}^{D}\mathbf{J}_{i,p}(\boldsymbol{\theta})\mathbf{C}_{i}
\mathbf{J}_{i,q}^{H}(\boldsymbol{\theta}) \ \
\text{for} \ \ p<q, \ \ (p,q) \in \{1, \ldots, M\}^2,
\end{equation}
where $\mathbf{C}_{i}$ is a known matrix describing the polarization state of the \textit{i}-th calibrator source while $\mathbf{J}_{i,p}(\boldsymbol{\theta})$ stands for all the perturbations along the full corresponding signal path (from the \textit{i}-th source to the \textit{p}-th antenna) and is referred as a $2 \times 2$ Jones matrix \cite{thompson2008interferometry}. The aim of calibration is to estimate the parameter vector $\boldsymbol{\theta}$.
We rewrite (\ref{ME}) as
\begin{equation}
\tilde{\mathbf{v}}_{pq}(\boldsymbol{\theta})=\mathrm{vec}\Big(\mathbf{V}_{pq}(\boldsymbol{\theta})\Big)
= \sum_{i=1}^{D}\mathbf{u}_{i,pq}(\boldsymbol{\theta})
\end{equation}
in which $\mathbf{u}_{i,pq}(\boldsymbol{\theta})=\left(\mathbf{J}^{\ast}_{i,q}(\boldsymbol{\theta})\otimes
\mathbf{J}_{i,p}(\boldsymbol{\theta})\right)\mathbf{c}_{i}$ and $\mathbf{c}_{i}=\mathrm{vec}(\mathbf{C}_{i})$. The noisy output observation vector of the full array is $\mathbf{x}=\left[\mathbf{v}^{T}_{12},
\mathbf{v}^{T}_{13},
\ldots,
\mathbf{v}^{T}_{(M-1)M}\right]^{T}$ \textit{s.t.} $\mathbf{v}_{pq}=\tilde{\mathbf{v}}_{pq}(\boldsymbol{\theta})+\mathbf{n}_{pq}$. We note $\mathbf{n}_{pq}$ the noise sample at a particular antenna pair which takes into account background Gaussian noise but also the presence of outliers.

To deal with non-Gaussianity of the noise, we assume a compound-Gaussian noise model since it includes a broad range of heavy-tailed distributions \cite{wang2006maximum,ollila2012complex}. Its expression is given by
\begin{equation}
\label{sirp}
\mathbf{n}_{pq} = \sqrt{\tau_{pq}} \  \boldsymbol{\mu}_{pq},
\end{equation} 
where $\tau_{pq}$ is the positive texture variable and $\boldsymbol{\mu}_{pq} \sim \mathcal{CN}(\mathbf{0},\boldsymbol{\Omega})$ is the complex speckle part.

\subsection{Robust estimation of Jones matrices}

The ML method is used to estimate iteratively parameters of interest $\boldsymbol{\theta}$, the covariance matrix $\boldsymbol{\Omega}$ and all realizations $\tau_{pq}$ for $\ p<q, \ \ (p,q) \in \{1, \ldots, M\}^2$ which are considered unknown and deterministic in the algorithm, leading to a relaxed version of the ML \cite{conte2002recursive}. Results are directly exposed in non-structured calibration algorithm (NSCA) and details can be found in \cite{ollier2016relaxed,ollierjournal}.
 Let us denote  $\mathbf{a}_{pq}(\boldsymbol{\theta}) = \mathbf{v}_{pq}- \tilde{\mathbf{v}}_{pq}(\boldsymbol{\theta})$, $B=\frac{M(M-1)}{2}$ is the number of antenna pairs and we impose $\mathrm{tr}\left\{\boldsymbol{\Omega}\right\}=1$ to remove scaling ambiguities \cite{zhang2016mimo}.
 Estimation of $\boldsymbol{\theta}$ in step 2 of NSCA can be performed numerically or thanks to expectation maximization (EM) \cite{mclachlan2007algorithm} and block coordinate descent (BCD) algorithms \cite{hong2015unified}. Such vector refers here to the entries of all $DM$ Jones matrices (non-structured case) \cite{yatawatta2009radio,nunhokeeghost2015}.

\begin{algorithm}
\SetAlgorithmName{NSCA}{}{} \caption{Non-structured calibration algorithm}
\SetKwInOut{input}{input} \SetKwInOut{output}{output}
\SetKwInOut{initialize}{initialize}
\input{$D$, $M$, $B$, $\{\mathbf{C}_{i}\}_{i=1,\hdots,D}$, $\mathbf{x}$}
\output{$\hat{\boldsymbol{\theta}}$}
\initialize{$\hat{\boldsymbol{\Omega}}$ $\leftarrow$ $\boldsymbol{\Omega}_{\mathrm{init}}$, $\{\hat{\boldsymbol{\tau}}_{pq}$ $\leftarrow$ $\boldsymbol{\tau}_{pq_{\mathrm{init}}}\}_{p<q,(p,q) \in \{1,\hdots,M\}^2} $}
\While
{stop criterion unreached}
{
\ShowLn $\hat{\boldsymbol{\theta }}=\argmin\limits_{\boldsymbol{\theta } }
\left\{\sum\limits_{pq}\frac{1}{\hat{\tau}_{pq}}\mathbf{a}_{pq}^{H}(\boldsymbol{\theta})
\hat{\boldsymbol{\Omega}}^{-1}\mathbf{a}_{pq}(\boldsymbol{\theta})\right\}$\\
\ShowLn  
$\hat{\boldsymbol{\Omega}}= \frac{4}{B}
\sum\limits_{pq} \frac{\mathbf{a}_{pq}(\hat{\boldsymbol{\theta}})
\mathbf{a}_{pq}^{H}(\hat{\boldsymbol{\theta}})}{\mathbf{a}_{pq}^{H}(\hat{\boldsymbol{\theta}})(\hat{\boldsymbol{\Omega}})^{-1}\mathbf{a}_{pq}(\hat{\boldsymbol{\theta}})}$ and $\hat{\boldsymbol{\Omega}}=\frac{\hat{\boldsymbol{\Omega}}}{\mathrm{tr}\left\{\hat{\boldsymbol{\Omega}}\right\}}$
\\
\ShowLn  $\{\hat{\tau}_{pq} = \frac{1}{4}
\mathbf{a}_{pq}^{H}(\hat{\boldsymbol{\theta}})
\hat{\boldsymbol{\Omega}}^{-1}
\mathbf{a}_{pq}(\hat{\boldsymbol{\theta}})\}_{p<q,(p,q) \in \{1,\hdots,M\}^2} $\\
}
\end{algorithm}

\section{Calibration in 3DC regime}

\subsection{3DC calibration regime}

We consider calibration of radio interferometry stations where antennas are enclosed in a compact array and direction dependent effects are dominant \cite{wijnholds2010calibration}. If we intend to calibrate the data at different frequencies, then a particular decomposition of the Jones matrix is \cite{noordam2010meqtrees,yatawatta2012reduced}
\begin{equation}
 \label{model_regime3}
\mathbf{J}^{[f]}_{i,p}(\boldsymbol{\theta}^{[f]}_{i,p})=\mathbf{G}_p(\mathbf{g}_p)\mathbf{H}^{[f]}_{i,p}\mathbf{Z}^{[f]}_{i,p}(\boldsymbol{\alpha}^{[f]}_i)\mathbf{F}^{[f]}_{i}(\vartheta^{[f]}_{i}) 
\end{equation}
for $i\in \{1, \hdots, D\}$, $p \in \{1, \hdots, M\}$, $f \in \mathcal{F}=\{f_1,\hdots,f_F\}$ and $\boldsymbol{\theta}^{[f]}_{i,p}=[\vartheta^{[f]}_{i},\mathbf{g}^T_p,\boldsymbol{\alpha}^{[f]^T}_i]^T$. Each individual Jones term and its corresponding effect is described in the following:


\subsubsection{Ionospheric effects}
  While travelling through a charged medium as the ionosphere, the incoming wave is affected by a phase delay due to refraction, and written as \cite{smirnov2011revisiting2}
  \begin{equation}
  \label{form_Z}
  \mathbf{Z}^{[f]}_{i,p}(\boldsymbol{\alpha}^{[f]}_i)=\exp\Big(j\varphi^{[f]}_{i,p}\Big)\mathbf{I}_2
  \end{equation} 
  where $\varphi^{[f]}_{i,p}=\eta^{[f]}_i u^{[f]}_p+\zeta^{[f]}_i v^{[f]}_p$ in which  $\boldsymbol{\alpha}^{[f]}_i=[\eta^{[f]}_i,\zeta^{[f]}_i]^T \propto f^{-2}$ \cite{cohen2009probing} stands for the apparent position shift in the source location and $\mathbf{r}^{[f]}_p=[u^{[f]}_p,v^{[f]}_p]^T$ is the known position vector of the \textit{p}-th antenna in wavelength units. Therefore, we deduce that $\varphi^{[f]}_{i,p} \propto f^{-1}$ \cite{van2009bayesian}.
Propagation through the upper part of the atmosphere also results in a rotation of the polarization plane, called Faraday rotation \cite{davies1990ionospheric},
\begin{equation}
\mathbf{F}^{[f]}_{i}(\vartheta^{[f]}_{i}) = \begin{bmatrix}
    \cos(\vartheta^{[f]}_{i}) & -\sin(\vartheta^{[f]}_{i}) \\
   \sin(\vartheta^{[f]}_{i}) & \cos(\vartheta^{[f]}_{i})
\end{bmatrix}
\end{equation}
where the Faraday rotation angle $\vartheta^{[f]}_i $  $\propto f^{-2}$ \cite{noordam2010meqtrees} is the same for all antennas due to the compact geometry of the array \cite{wijnholds2010calibration}.


\subsubsection{Instrumental effects}
Receiver electronics introduce direction and frequency independent effects \cite{noordam2010meqtrees}, leading to a complex sensor gain matrix, noted $\mathbf{G}_p(\mathbf{g}_p)=\mathrm{diag}\{\mathbf{g}_p\}$.

Finally, let us note that $\mathbf{H}^{[f]}_{i,p}$ is a known matrix given by electromagnetic modeling and a priori information (calibrator sources and antenna positions) \cite{thompson2008interferometry}. Let us recall that the combined effect of all individual perturbations along a particular signal path is represented by a global Jones matrix, as shown in (\ref{model_regime3}).

\subsection{Mono-frequency case}

In 3DC regime, calibration amounts to estimate the physical parameters $\boldsymbol{\theta}_{i,p}$  which appear in the structured Jones matrices (\ref{model_regime3}). In the mono-frequency case, we use an alternating least squares approach to estimate unknowns sequentially \cite{wijnholds2010fish}, by optimizing a cost function \textit{w.r.t.} one parameter while fixing the others. The global frame of the structured calibration algorithm (SCA) is exposed hereafter, for one given $i$ and $p$, and details can be found in \cite{ollierjournal}. Let us note that $\hat{\boldsymbol{\varphi}}_i=[\hat{\varphi}_{i,1},\hdots,\hat{\varphi}_{i,M}]^T$ and $\boldsymbol{\Lambda}= [\mathbf{r}_1,  \hdots  ,\mathbf{r}_M]$.

 \begin{algorithm}
\SetAlgorithmName{SCA}{}{} \caption{Structured calibration algorithm}
\SetKwInOut{input}{input} \SetKwInOut{output}{output}
\SetKwInOut{initialize}{initialize}
\input{$D$, $M$, $B$, $\mathbf{C}_{i}$, $\mathbf{x}$, $\hat{\mathbf{J}}_{i,p}$ as output of NSCA}
\output{$\hat{\boldsymbol{\theta}}_{i,p}$}
\initialize{$\hat{\boldsymbol{\theta}}_{i,p} \leftarrow \boldsymbol{\theta}_{{i,p}_{\mathrm{init}}}$}
\While
{stop criterion unreached}
{ \ShowLn  $\hat{\vartheta}_i= \argmin_{\vartheta_i} \sum_{p=1}^M ||\hat{\mathbf{J}}_{i,p}-\mathbf{G}_p\mathbf{H}_{i,p}\mathbf{Z}_{i,p}\mathbf{F}_{i}(\vartheta_{i})||^2_{F}$ \\
\ShowLn  $\hat{\mathbf{g}}_p= \argmin_{\mathbf{g}_p}\sum_{i=1}^D  ||\hat{\mathbf{J}}_{i,p}-\mathbf{G}_p(\mathbf{g}_p)\mathbf{H}_{i,p}\mathbf{Z}_{i,p}\mathbf{F}_{i}||^2_{F}$\\
\ShowLn $\hat{\varphi}_{i,p}= \argmin_{\varphi_{i,p}}||\hat{\mathbf{J}}_{i,p}-\mathbf{G}_p\mathbf{H}_{i,p}\mathbf{Z}_{i,p}(\varphi_{i,p})\mathbf{F}_{i}||^2_{F}$ \\
\ShowLn $ \hat{\boldsymbol{\alpha}}_i^T=
 \frac{\hat{\boldsymbol{\varphi}}_i^T\boldsymbol{\Lambda}^H \begin{bmatrix}
  \sum_{p=1}^M v_p^2 & -\sum_{p=1}^M u_p v_p\\
-\sum_{p=1}^M v_p u_p &  \sum_{p=1}^M u_p^2
  \end{bmatrix}}{\sum_{p=1}^M u_p^2 \sum_{p=1}^M v_p^2 - (\sum_{p=1}^M u_p v_p)^2}$ \\
}
\end{algorithm}

\subsection{Multi-frequency calibration algorithm}

The aim of calibration in a multi-frequency scenario is to estimate the parameter vector of interest
 $\boldsymbol{\epsilon}=[\boldsymbol{\epsilon}^{[f_1]^T}, \hdots,\boldsymbol{\epsilon}^{[f_F]^T}, \mathbf{g}^T]^T$
  where $\boldsymbol{\epsilon}^{[f]}=[\vartheta^{[f]}_1,\hdots,\vartheta^{[f]}_D,\boldsymbol{\alpha}^{[f]^T}_1,\hdots,\boldsymbol{\alpha}^{[f]^T}_D]^T$ and $\mathbf{g}=[\mathbf{g}_1^T,\hdots,\mathbf{g}_M^T]^T$. To do so, we introduce the following cost function
   \begin{equation}
   \label{cost_original}
l^{[f]}(\boldsymbol{\epsilon}^{[f]} )  = \sum_{i=1}^D l_{i}^{[f]}(\boldsymbol{\epsilon}_{i}^{[f]} ) 
\end{equation}
in which  
$l_{i}^{[f]}(\boldsymbol{\epsilon}_{i}^{[f]} )= \sum\limits_{p=1}^M || \hat{\mathbf{J}}^{[f]}_{i,p}-\mathbf{G}_p\mathbf{H}^{[f]}_{i,p}\mathbf{Z}^{[f]}_{i_{p}}(\boldsymbol{\alpha}^{[f]}_i)\mathbf{F}_i^{[f]}(\vartheta_{i}^{[f]})||^2_{F}$
with $\boldsymbol{\epsilon}^{[f]}_i=[\vartheta^{[f]}_i,\boldsymbol{\alpha}^{[f]^T}_i]^T$. Prior information on $\hat{\mathbf{J}}^{[f]}_{i,p}$ for $i \in \{1, \hdots, D\}$, $p \in\{1, \hdots, M\}$ and $f \in \mathcal{F}$ is provided by the output of NSCA.

We wish to distributedly solve the following constrained optimization problem thanks to a network of agents
   \begin{align}
   \label{constrain}
   \{\hat{\boldsymbol{\epsilon}}^{[f]}\}_{f \in \mathcal{F}}, \hat{\mathbf{z}} & = \argmin_{\boldsymbol{\epsilon}^{[f_1]},\hdots,\boldsymbol{\epsilon}^{[f_F]},\mathbf{z}} \sum_{f \in \mathcal{F}} l^{[f]}(\boldsymbol{\epsilon}^{[f]} ) \\&
   \textit{s.t.} \ \  \boldsymbol{\epsilon}^{[f]}_i= \mathbf{B}^{[f]}\mathbf{z}_i, \ \ i \in \{1,\ldots,D\},  f \in \mathcal{F}
   \end{align}
 where $\mathbf{B}^{[f]}=\frac{1}{f^2}\mathbf{I}_{3}$ is the known frequency model, $\mathbf{z}_i$ is an unknown associated global variable, independent \textit{w.r.t.} frequency and shared by all agents, and $\mathbf{z}=[\mathbf{z}_1^T,\hdots,\mathbf{z}_D^T]^T$. 
To solve (\ref{constrain}), we use a consensus optimization scheme as in the ADMM procedure \cite{boyd2011distributed}. Instead of considering the original objective function (\ref{cost_original}), we study the following augmented Lagrangian
   \begin{equation}
   L(\boldsymbol{\epsilon}^{[f_1]},\hdots,\boldsymbol{\epsilon}^{[f_F]},\mathbf{z},\mathbf{y}^{[f_1]},\hdots,\mathbf{y}^{[f_F]})  
   = \sum_{f \in \mathcal{F}}  \sum\limits_{i=1}^D
L_{i}^{[f]}\left(\boldsymbol{\epsilon}_{i}^{[f]},\mathbf{z}_i,\mathbf{y}_{i}^{[f]}\right)
    \end{equation}
where $L_{i}^{[f]}\left(\boldsymbol{\epsilon}_{i}^{[f]},\mathbf{z}_i,\mathbf{y}_{i}^{[f]}\right)= l_{i}^{[f]}\left(\boldsymbol{\epsilon}_{i}^{[f]} \right)+  h_{i}^{[f]}\left(\boldsymbol{\epsilon}_{i}^{[f]},\mathbf{z}_i,\mathbf{y}_{i}^{[f]}\right)$ and 
\begin{equation}
h_{i}^{[f]}\left(\boldsymbol{\epsilon}_{i}^{[f]},\mathbf{z}_{i},\mathbf{y}_{i}^{[f]}\right)=\mathbf{y}_{i}^{[f]^T}\left(\boldsymbol{\epsilon}_{i}^{[f]}-\mathbf{B}^{[f]}\mathbf{z}_{i} \right)+\frac{\rho}{2}||\boldsymbol{\epsilon}_{i}^{[f]}-\mathbf{B}^{[f]}\mathbf{z}_{i}||^2_2.
\end{equation}
We note $\mathbf{y}^{[f]}=[\mathbf{y}^{[f]^T}_1,\hdots,\mathbf{y}^{[f]^T}_D]^T$ the associated Lagrange parameters (or dual variables) and $\rho > 0$ a penalty factor.  We notice separability of the Lagrangian  \textit{w.r.t.} source direction but above all, separability \textit{w.r.t.} frequency, meaning that each agent solves a subproblem locally at a given frequency.
The ADMM consists in updating sequentially the three following quantities:
\begin{align}
 \label{step1}
  \bullet  \left(\hat{\boldsymbol{\epsilon}}_{i}^{[f]}\right)^{t+1} & =  \argmin_{\boldsymbol{\epsilon}_{i}^{[f]}}L_{i}^{[f]}\left(\boldsymbol{\epsilon}_{i}^{[f]},\left(\hat{\mathbf{z}}_{i}\right)^{t},\left(\hat{\mathbf{y}}_{i}^{[f]}\right)^{t}\right) \\ &
    \nonumber    
    \ \text{performed locally by each agent for $i \in \{1,...,D\}$}
      \end{align}
      \begin{align}
       \label{step2}
  \bullet   \left(\hat{\mathbf{z}}_{i}\right)^{t+1} & =\argmin_{\mathbf{z}_{i}} \sum_{f \in \mathcal{F}} L_{i}^{[f]}\left(\left(\hat{\boldsymbol{\epsilon}}_{i}^{[f]}\right)^{t+1},\mathbf{z}_{i},\left(\hat{\mathbf{y}}_{i}^{[f]}\right)^{t}\right) \\& 
      \nonumber    
    \ \text{performed globally for $i \in \{1,...,D\}$}
      \end{align}
     \begin{align}
      \label{step3}
   \bullet  \left(\hat{\mathbf{y}}_{i}^{[f]}\right)^{t+1}& =\left(\hat{\mathbf{y}}_{i}^{[f]}\right)^{t}+ \rho\left(\left(\hat{\boldsymbol{\epsilon}}_{i}^{[f]}\right)^{t+1} -  \mathbf{B}^{[f]}\left(\hat{\mathbf{z}}_{i}\right)^{t+1}\right) \\& 
        \nonumber    
    \ \text{performed locally by each agent for $i \in \{1,...,D\}$}
      \end{align}
      where $t$ is the iteration counter.
 Minimization (\ref{step2}) needs access to local solutions from all agents, \textit{i.e.}, at all frequencies, and leads to the following  closed-form expression \cite{yatawatta2015distributed,boyd2011distributed},
       \begin{equation}
       \label{estim_c}
      \hat{\mathbf{z}}_i=\left(\sum_{f \in \mathcal{F}}\rho\mathbf{B}^{[f]^T}\mathbf{B}^{[f]}\right)^{-1}\left(\sum_{f \in \mathcal{F}}\mathbf{B}^{[f]^T}(\mathbf{y}^{[f]}_i+\rho\boldsymbol{\epsilon}^{[f]}_i)\right).
       \end{equation}      
Minimization (\ref{step1}) is addressed iteratively. To this end, we compute the gradient of $L_i^{[f]}(\boldsymbol{\epsilon}_i^{[f]},\mathbf{z}_i,\mathbf{y}_i^{[f]} )$ \textit{w.r.t.} $\boldsymbol{\epsilon}_i^{[f]}$, which induces
       \begin{equation}
       \label{deriv1}
       \frac{\partial L_i^{[f]}(\boldsymbol{\epsilon}_i^{[f]},\mathbf{z}_i,\mathbf{y}_i^{[f]} )}{\partial\vartheta_i^{[f]}}= [\mathbf{y}_i^{[f]}]_1 + \rho(\vartheta_i^{[f]}-\frac{1}{f^2}[\mathbf{z}_i]_1)+
     \sum_{p=1}^M \mathrm{tr}\left\{\mathbf{S}^{[f]}_{i,p}+\mathbf{S}^{[f]^H}_{i,p}
 \right\} 
       \end{equation}
       where $\mathbf{S}^{[f]}_{i,p}=-\mathbf{G}_p\mathbf{H}^{[f]}_{i,p}\mathbf{Z}^{[f]}_{i,p}\frac{\partial \mathbf{F}_i^{[f]}(\vartheta_i^{[f]})}{\partial\vartheta_i^{[f]}}\hat{\mathbf{J}}^{[f]^H}_{i,p}$, and
        \begin{equation}
        \label{deriv2}
       \frac{\partial L_i^{[f]}(\boldsymbol{\epsilon}_i^{[f]},\mathbf{z}_i,\mathbf{y}_i^{[f]} )}{\partial\eta_i^{[f]}}= 
[\mathbf{y}_i^{[f]}]_2 + \rho(\eta_i^{[f]}-\frac{1}{f^2}[\mathbf{z}_i]_2)+
\sum_{p=1}^M \mathrm{tr}\left\{\mathbf{D}^{[f]}_{i,p} +\mathbf{D}^{[f]^H}_{i,p} \right\} 
        \end{equation}
        where $\mathbf{D}^{[f]}_{i,p}=
 j u^{[f]}_{p} \mathbf{Z}^{[f]^{\ast}}_{i,p}
    \mathbf{M}^{[f]}_{i,p}
 $ and $\mathbf{M}^{[f]}_{i,p}=\hat{\mathbf{J}}^{[f]}_{i,p}\mathbf{F}^{[f]^T}_{i}\mathbf{H}^{[f]^H}_{i,p}\mathbf{G}_p^H$. Likewise, we have
    \begin{equation}
    \label{deriv3}
       \frac{\partial L_i^{[f]}(\boldsymbol{\epsilon}_i^{[f]},\mathbf{z}_i,\mathbf{y}_i^{[f]} )}{\partial\zeta_i^{[f]}}= 
[\mathbf{y}_i^{[f]}]_3 + \rho(\zeta_i^{[f]}-\frac{1}{f^2}[\mathbf{z}_i]_3)+
\sum_{p=1}^M \mathrm{tr}\left\{\mathbf{V}^{[f]}_{i,p} +\mathbf{V}^{[f]^H}_{i,p} \right\} 
        \end{equation}
        where $\mathbf{V}^{[f]}_{i,p}=
 j v^{[f]}_{p} \mathbf{Z}^{[f]^{\ast}}_{i,p}
   \mathbf{M}^{[f]}_{i,p}  $.
Using \crefrange{deriv1}{deriv3}, we obtain $\hat{\boldsymbol{\epsilon}}_i^{[f]}$ with a root-finding algorithm or thanks to standard numerical optimization tools as Newton or gradient descent-type algorithm \cite{nocedal2006numerical}.

Estimation of the gains $\mathbf{g}_p$ is done as
 minimization of the following least squares cost function 
\begin{equation}
\label{gain_cost}
  \kappa(\mathbf{g}_p)=\sum_{f \in \mathcal{F}}  \sum_{i=1}^D ||\hat{\mathbf{J}}^{[f]}_{i,p}-\mathbf{G}_p(\mathbf{g}_p)\mathbf{H}^{[f]}_{i,p}\mathbf{Z}^{[f]}_{i,p}\mathbf{F}_i^{[f]}||^2_{F}
\end{equation}
leading to the following estimate of each complex gain element for $k \in \{1,2\}$
 \begin{equation}
  \label{estim_gp_final}
[\hat{\mathbf{g}}_p]_k=\Big(\sum_{f \in \mathcal{F}}\sum_{i=1}^D[\mathbf{W}^{[f]^{\ast}}_{i,p}]_{k,k}\Big)^{-1} \sum_{f \in \mathcal{F}}\sum_{i=1}^D[\mathbf{X}^{[f]^{\ast}}_{i,p}]_{k,k}
\end{equation}
where $\mathbf{X}^{[f]}_{i,p}=\mathbf{R}^{[f]}_{i,p}\hat{\mathbf{J}}^{[f]^H}_{i,p}$ and $\mathbf{W}^{[f]}_{i,p}=\mathbf{R}^{[f]}_{i,p}\mathbf{R}^{[f]^H}_{i,p}$ in which $\mathbf{R}^{[f]}_{i,p}=\mathbf{H}^{[f]}_{i,p}\mathbf{Z}^{[f]}_{i,p}\mathbf{F}^{[f]}_{i}$.
The scheme of the proposed multi-frequency structured calibration algorithm (MSCA) is described below.

\begin{algorithm}
\SetAlgorithmName{MSCA}{}{} \caption{Multi-frequency structured calibration algorithm}
\SetKwInOut{input}{input} \SetKwInOut{output}{output}
\SetKwInOut{initialize}{initialize}
\input{$D$, $M$, $F$, $\mathbf{r}_p^{[f]}$, $\mathbf{H}^{[f]}_{i,p}$, $\hat{\mathbf{J}}^{[f]}_{i,p}$ as output of NSCA for $i \in \{1, \hdots, D\}$, $p \in \{1, \hdots, M\}$ and $f \in \mathcal{F}$}
\output{$\hat{\boldsymbol{\epsilon}}$}
\initialize{$\hat{ \boldsymbol{\epsilon}}$ $\leftarrow$ $ \boldsymbol{\epsilon}_{\mathrm{init}}$, $\hat{\mathbf{z}}$ $\leftarrow$ $\mathbf{z}_{\mathrm{init}}$, $\{\hat{\mathbf{y}}^{[f]}$ $\leftarrow$ $\mathbf{y}^{[f]}_{\mathrm{init}}\}_{f\in \mathcal{F}}$}
\While
{stop criterion unreached}
{
\While 
{stop criterion unreached}
{
\While   
{ stop criterion unreached}{
\ShowLn Obtain $\{\hat{\vartheta}_i^{[f]}\}_{i =1,\hdots,D} $ locally with (\ref{deriv1})\\
\ShowLn Obtain $\{\hat{\eta}_i^{[f]}\}_{i =1,\hdots,D}$ locally with (\ref{deriv2})\\
\ShowLn Obtain $\{\hat{\zeta}_i^{[f]}\}_{i =1,\hdots,D}$ locally with (\ref{deriv3})\\
}
\ShowLn Obtain $\{\hat{\mathbf{z}}_i\}_{i = 1,\hdots,D}$ globally with (\ref{estim_c})\\
\ShowLn Obtain $\{\hat{\mathbf{y}}_i^{[f]}\}_{i=1,\hdots,D}$ locally with (\ref{step3})\\
}
\ShowLn Obtain $\{\hat{\mathbf{g}}_p\}_{p = 1,\hdots,M}$ with (\ref{estim_gp_final})\\
}
\end{algorithm}

\section{NUMERICAL SIMULATIONS}

\begin{figure}[t!] 
  \centering
  \centerline{\includegraphics[width=0.45\textwidth,height=0.30\textwidth]{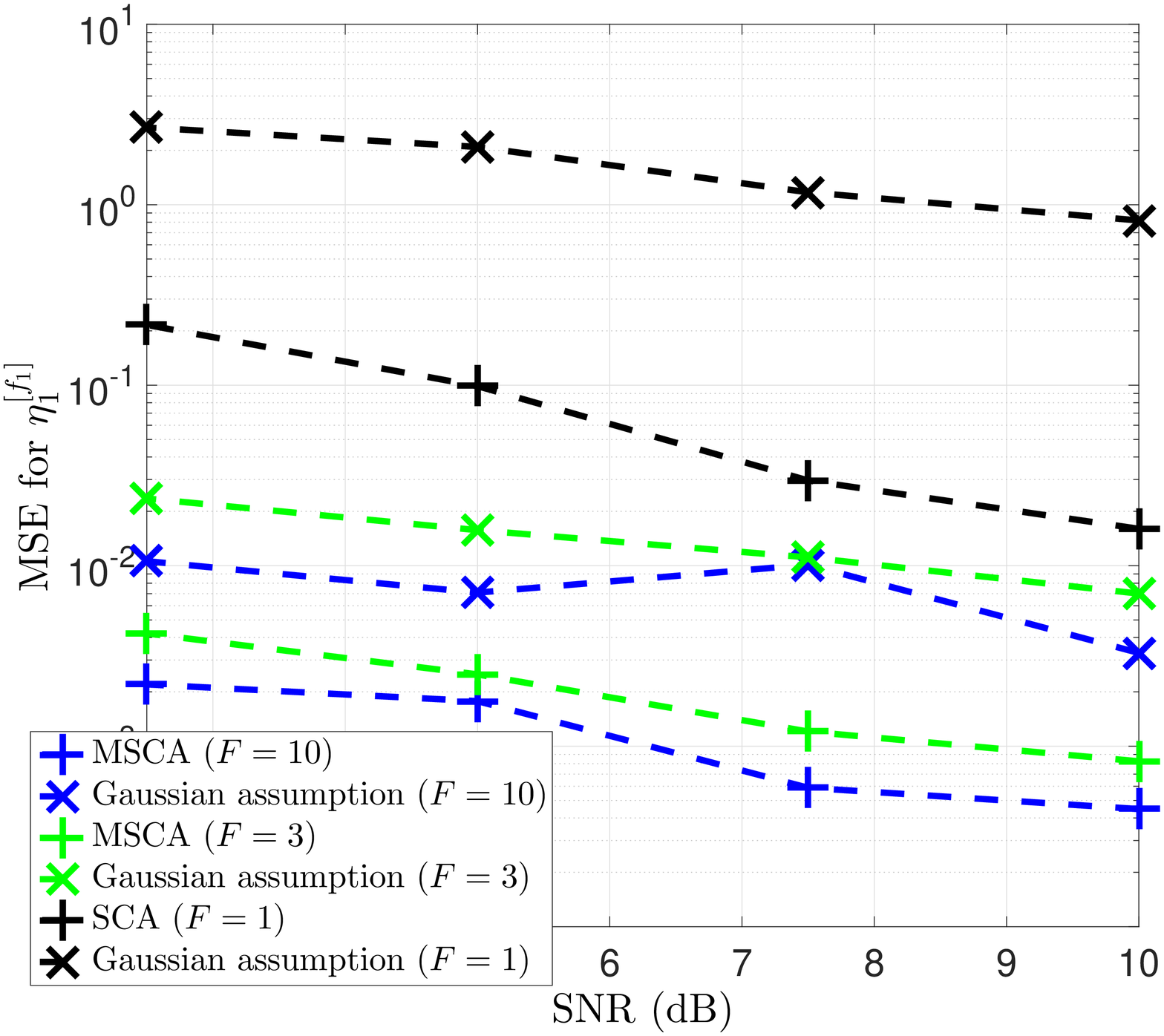}}
\caption{MSE of $\eta^{[f_1]}_1$ vs. SNR, for $D=2$ bright signal sources, $M=8$ antennas and $4$ weak outlier sources.} \label{fig:res1}

\end{figure}

\begin{figure}[t!]
 \centering
\subfigure[]{\label{fig:res2-a}\includegraphics[width=2.5cm, height=2.5cm]{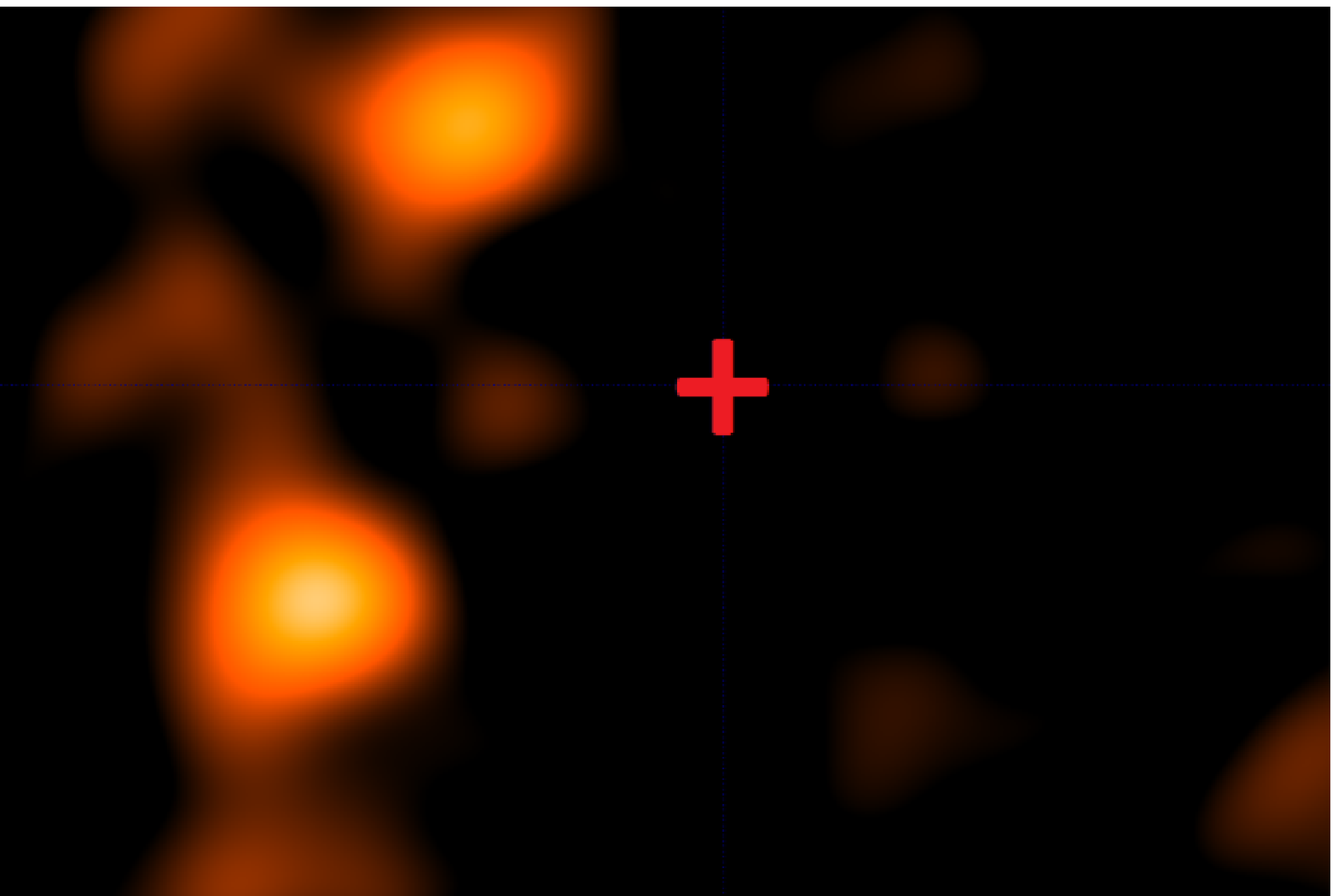}} \hspace{0.5pt} %
\subfigure[]{\label{fig:res2-b}\includegraphics[width=2.5cm, height=2.5cm]{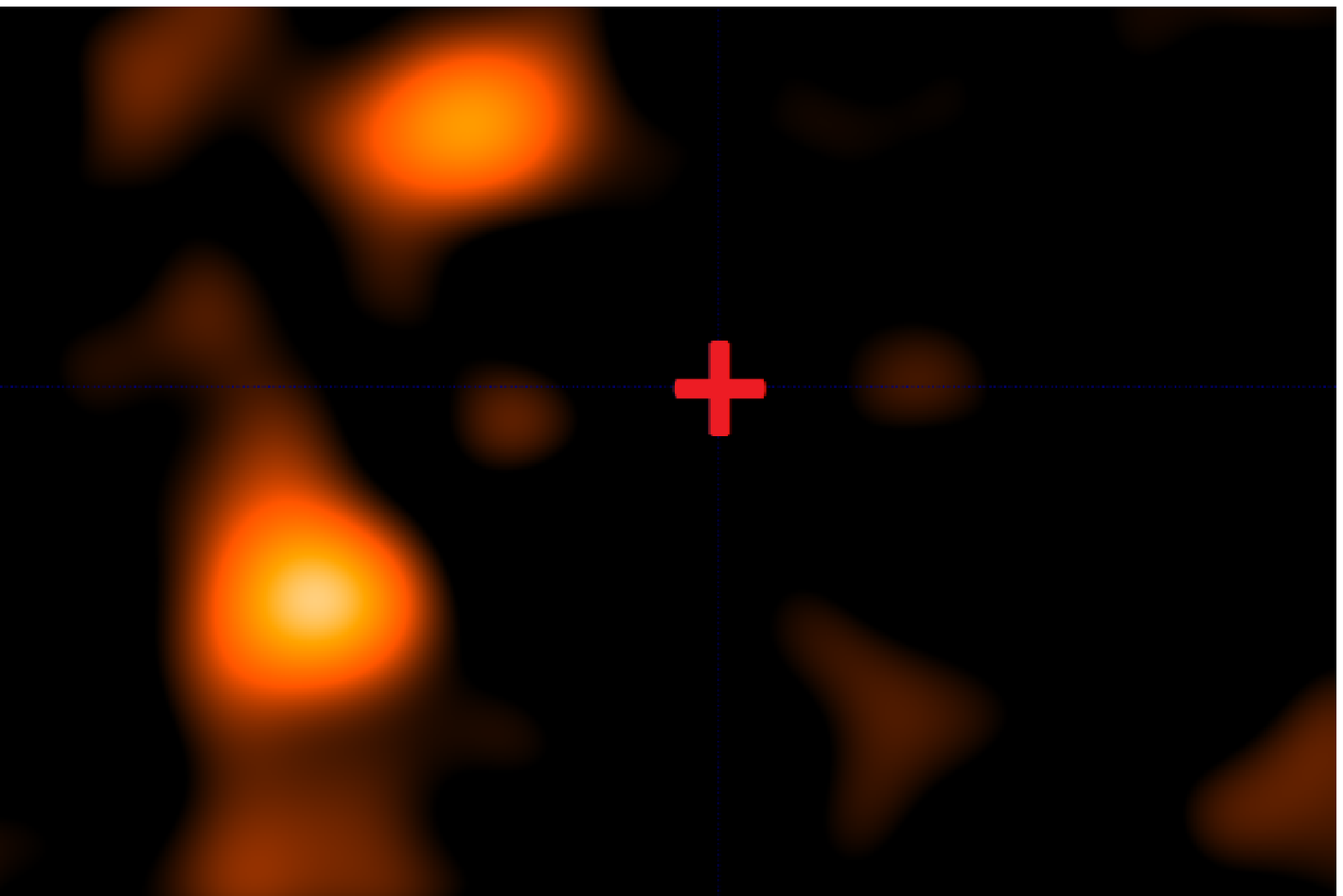}} 
\hspace{0.5pt} %
\subfigure[]{\label{fig:res2-c}\includegraphics[width=2.5cm, height=2.5cm]{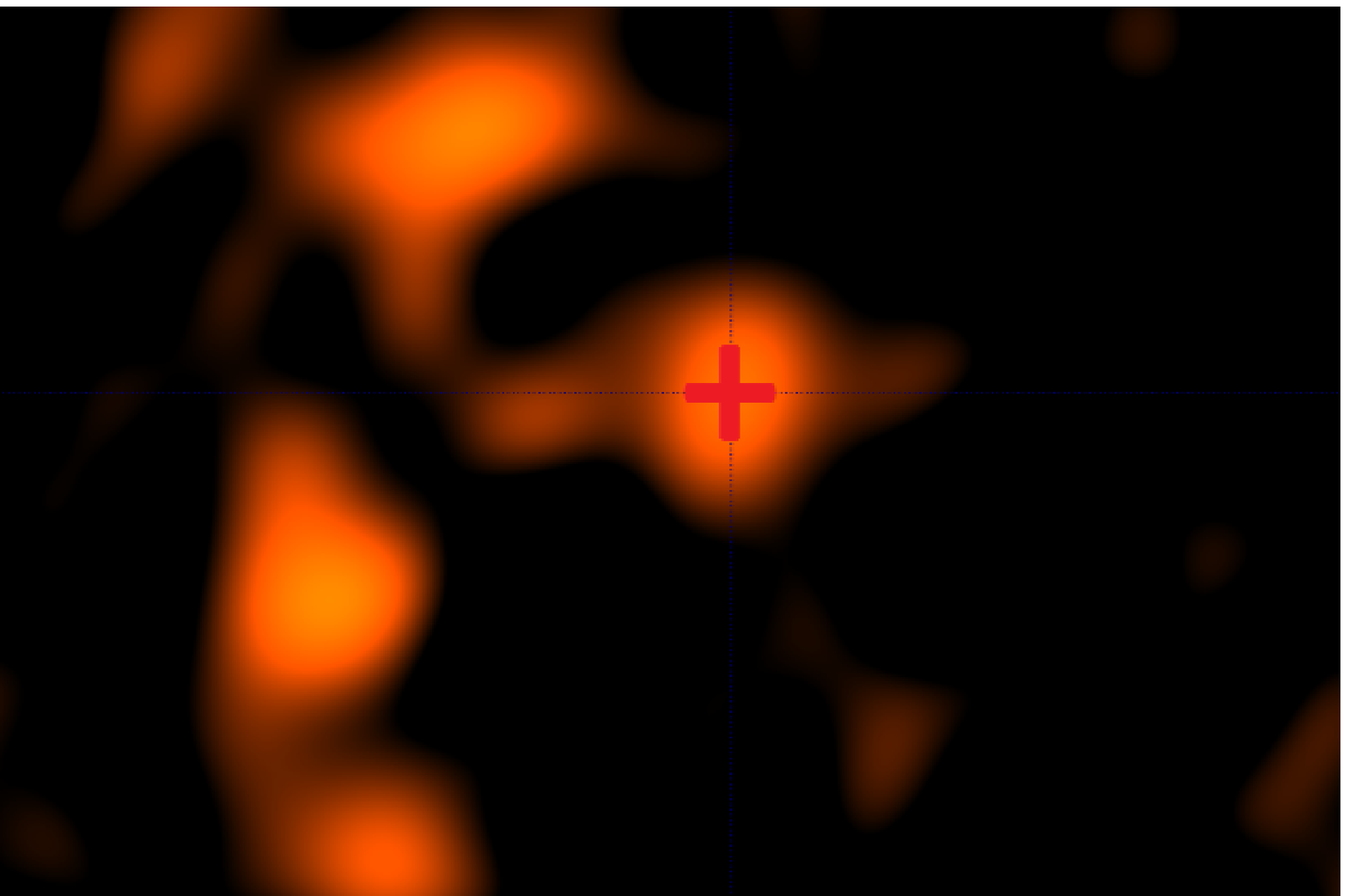}} 
\caption{\label{fig:res2} Corrected residual for (a) ideal case, (b) MSCA and (c) Meqtrees solver at $895$ MHz. The red cross corresponds to the location of the calibrator source. }
\end{figure}

\begin{figure}[t!] 
  \centering
  \centerline{\includegraphics[width=0.45\textwidth,height=0.30\textwidth]{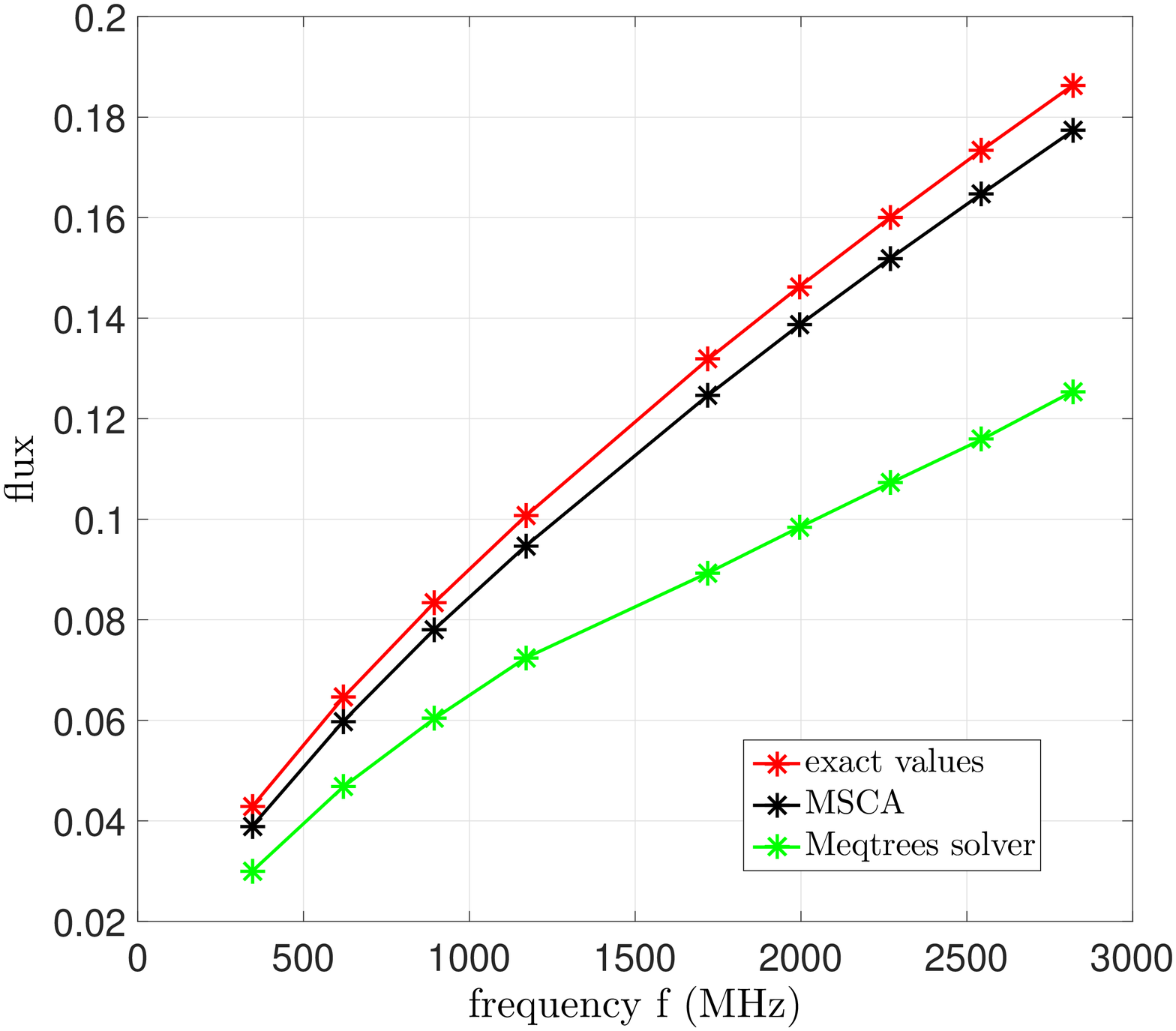}}
\caption{Recovered flux of one weak outlier source after calibration for $F=9$.} \label{fig:res3}

\end{figure}

In this section, we compare the performance of the proposed MSCA with the mono-frequency case, \textit{i.e.}, SCA in which frequency diversity is not taken into account. We recall that radio astronomy observations are affected by the presence of outliers. Thus, we also compare our robust approach with an algorithm based on a classical Gaussian noise assumption \cite{yatawatta2009radio}, which amounts to solve a non-linear least squares problem. First, in Fig. \ref{fig:res1}, we plot the mean square error (MSE)  of $\eta^{[f_1]}_1$ as a function of the signal-to-noise ratio (SNR), the behavior being the same for any other parameter of $\boldsymbol{\epsilon}$. We compare the estimation performance for different number of frequencies $F$ and notice better statistical performance when multi-frequency robust calibration is performed.  Robust calibration based on the Student's t \cite{yatawatta2014robust} is not exposed in the simulations due to a different model which is not adapted to the 3DC regime.\\
\indent
In the following figures, we use Meqtrees \cite{noordam2010meqtrees} to generate the data model, the observations and compare its least squares solver to MSCA. Here, we choose to correct for Faraday rotation matrices, which are the only introduced perturbations in the observations. We consider $M=7$ antennas (KAT-7 instrument), $D=1$, $D'=16$ weak realistic background sources taken from the SUMSS survey using a spectral index of $0.7$. The full duration of the observation is $12$ hours, for $60$ seconds integration time per data sample. 
After calibration and subtraction of the bright calibrator source, a dirty image, namely the corrected residual, is constructed with Meqtrees using \textit{lwimager}.  Fig. \ref{fig:res2} gives the corrected residual image at $895$ MHz in a small area surrounding the calibrator, whose position corresponds to the red cross.  Fig. \ref{fig:res3} gives the recovered flux  for one of the $D’$ weak outlier sources. Therefore, we notice better flux estimation of weak background sources and better calibrator removal using joint frequency dependent calibration with MSCA compared to a frequency independent calibration.



%

\section{CONCLUSION}

This paper introduces a robust ML based calibration technique in the context of radio interferometry. Robustness is addressed thanks to a compound-Gaussian noise modeling and a particular scenario is studied, \textit{i.e.}, the 3DC calibration regime. Variation of parameters \textit{w.r.t.} frequency imposes additional constraints which are considered in an ADMM-based distributed algorithm.  We show in the simulations the advantages of the proposed algorithm in regards to standard mono-frequency and/or non-robust scenario.

%
\bibliographystyle{IEEEtran}
\bibliography{biblio}

\end{document}